\documentstyle[11pt,newpasp,twoside,epsf]{article}

\begin{document}

\title{Star Formation and Structure Formation at $1\la z\la 4$}
\author{Kurt Adelberger}
\affil{Palomar Observatory, Caltech 105--24, Pasadena, CA 91125}

\begin{abstract}
The advent of 8m-class telescopes has made galaxies
at $1\la z\la 4$ relatively easy to detect and study.
This is a brief and incomplete review of some of the
recent results to emerge from surveys at these redshifts.
After describing different strategies for
finding galaxies at $z\ga 1$, and the differences
(and similarities) in the resulting galaxy
samples, I summarize what is known about the
spatial clustering of star-forming galaxies
at $z\ga 1$.  Optically selected galaxies are
the main focus of this review, but in the final
section I discuss the connection between
optical and sub-mm samples, and argue that
the majority of the $850\mu$m background
may have been produced by known optically selected
populations at high redshift.
Among the new results presented are the dust-corrected
luminosity function of Lyman-break galaxies at $z\sim 3$,
the estimated contribution to the $850\mu$m background
from optically selected galaxies at $1\la z\la 5$,
revised estimates of the spatial clustering strength
of Lyman-break galaxies at $z\sim 3$, and an
estimate of the clustering strength of star-forming
galaxies at $z\simeq 1$ derived from a new spectroscopic
sample of $\sim 800$ galaxies with $\langle z\rangle = 1.0$,
$\sigma_z = 0.1$.

\end{abstract}


\section{Introduction}
An ambitious goal in cosmology is to understand how
the universe evolved from its presumed beginning
in the Big Bang to the familiar collection
of stars and galaxies that we observe around us today.
The last decade has seen tremendous progress in understanding
the large role that gravitational instability almost certainly played.
Although we still do not have complete analytic understanding, 
reasonable analytic approximations for the growth of
gravitationally driven perturbations are now known, and sophisticated N-body
simulators and simulations are freely available for obtaining
more precise or detailed information.  Remarkable progress has also
been made in observationally constraining the initial conditions
that are required as input to the simulations
or approximations.  The available data
appear largely consistent with the idea that primordial
fluctuations were Gaussian (e.g. Bromley \& Tegmark 1999)
with a power-spectrum similar to that of an adiabatic
$\Lambda$CDM model over $\ga 5$ orders of magnitude
in spatial scale (e.g. numerous recent results from
observations of the cosmic microwave background,
reviewed most recently by Scott 2000; White, Efstathiou, \& Frenk 1993; 
Croft et al. 1999).

But gravitational instability is only half---the easy half!---of the story.
It alone cannot tell us how or when the stars that populate the
universe today were formed.
Presumably stars began to form within overdensities
in the matter distribution as these overdensities slowly evolved from
small ripples in the initial conditions into
the large collapsed objects of today,
but modeling this has proved tremendously difficult.  
We cannot easily
model the formation of a single star (see Abel's contribution
to these proceedings), let
alone the ten billion stars in a typical galaxy.
Even the most sophisticated theoretical treatments
of galaxy formation rely on simplified ``recipes''
for associating the formation of stars with the
gravitationally driven growth
of perturbations in the underlying matter
distribution.  The adopted recipes for star formation,
although physically plausible, are by far the
most uncertain component in theoretical treatments
of galaxy formation.  We will need to check
them through observations of star-forming galaxies
at high-redshift before we can be confident
that our understanding of galaxy formation is reasonably correct.
These observations,
and their implications, are the subject of my talk.

\section{Finding Galaxies at $z\ga 1$}
In the past 5 years several techniques have been shown 
effective for finding galaxies at $z\ga 1$.  I don't
have space to list them all; a partial list would include
deep optical magnitude limited
surveys (e.g. Cohen's contribution to these proceedings), 
narrow band surveys (e.g. Hu, Cowie, \& McMahon 1998), targeted surveys
around known AGN at $z\ga 1$ (e.g. Hall \& Green 1998, Djorgovski et al. 1999),
$850\mu$m surveys (e.g. Ivison et al. 2000), and color-selected
surveys (e.g. Steidel et al. 1999, Adelberger et al. 2000).  
Different selection techniques have different advantages
and are optimized for answering different questions.
Color-selected surveys, which detect numerous galaxies
over large and (hopefully) representative volumes,
are especially well suited for studying large scale
structure at high redshift.  They will be the main
focus of this review.

In color-selected surveys, spectra are obtained only for
objects with broad-band colors indicating that
they are likely to lie at a given redshift.
The left panel of Figure 1 illustrates why galaxies
at certain redshifts have distinctive broad-band colors.
The right panel shows spectroscopic redshifts 
for galaxies satisfying various simple color selection criteria,
demonstrating that color selection is a reasonably
effective way of finding galaxies at
a range of redshifts $z\ga 1$.
At $z\sim 1$ these galaxies were selected by exploiting the Balmer-break (Adelberger et al. 2000),
at $z\sim 3$ and $z\sim 4$ by exploiting the Lyman-break (Steidel et al. 1999),
and at $z\sim 2.2$ by an approach similar to the ``UV drop in'' technique
described by Roukema in these proceedings.
The data in Figure 1 represent only our own efforts;
many more galaxies at similar (and higher) redshifts have been found by other
groups with a variety of techniques.
    
\begin{figure}
\plottwo{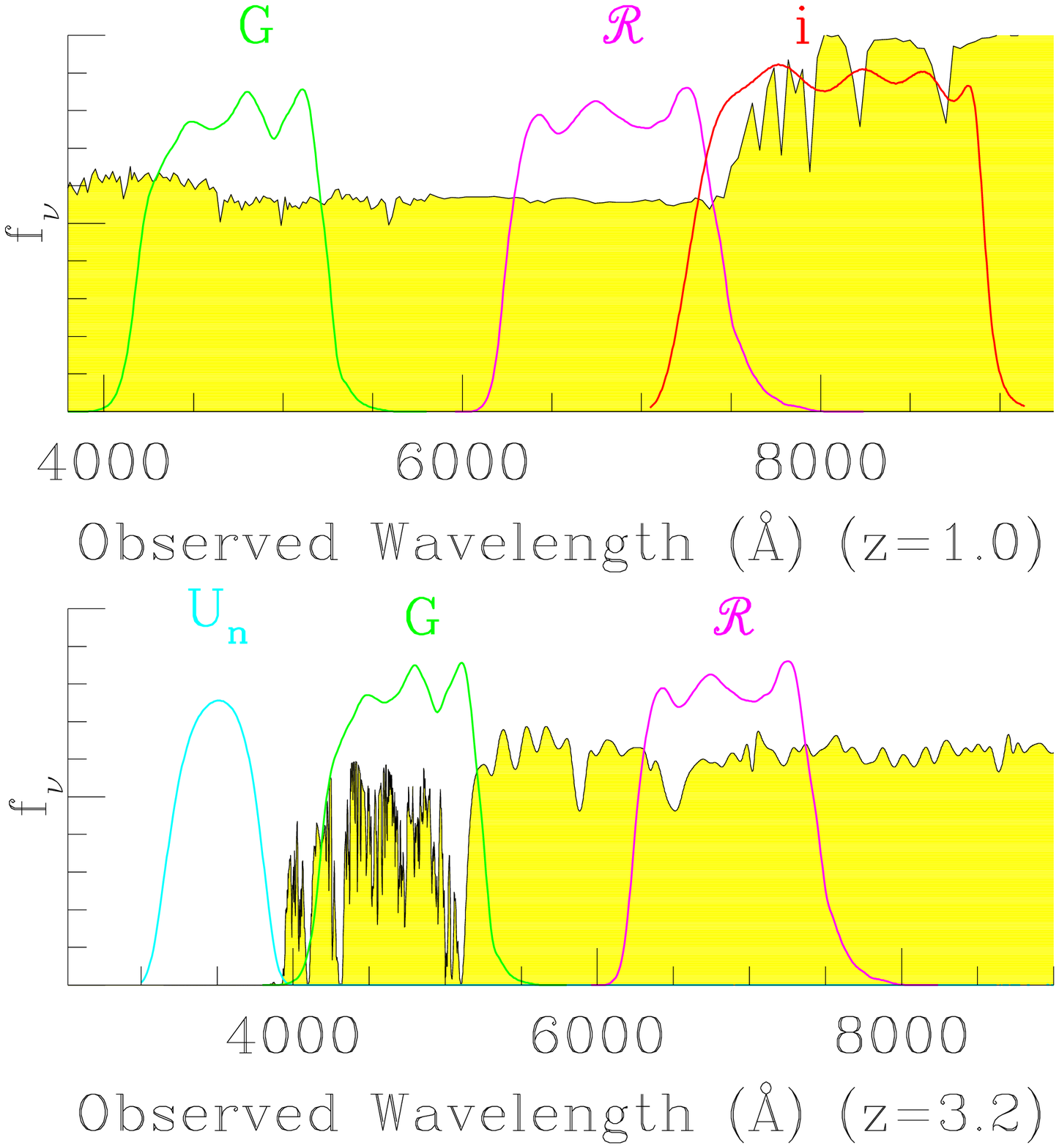}{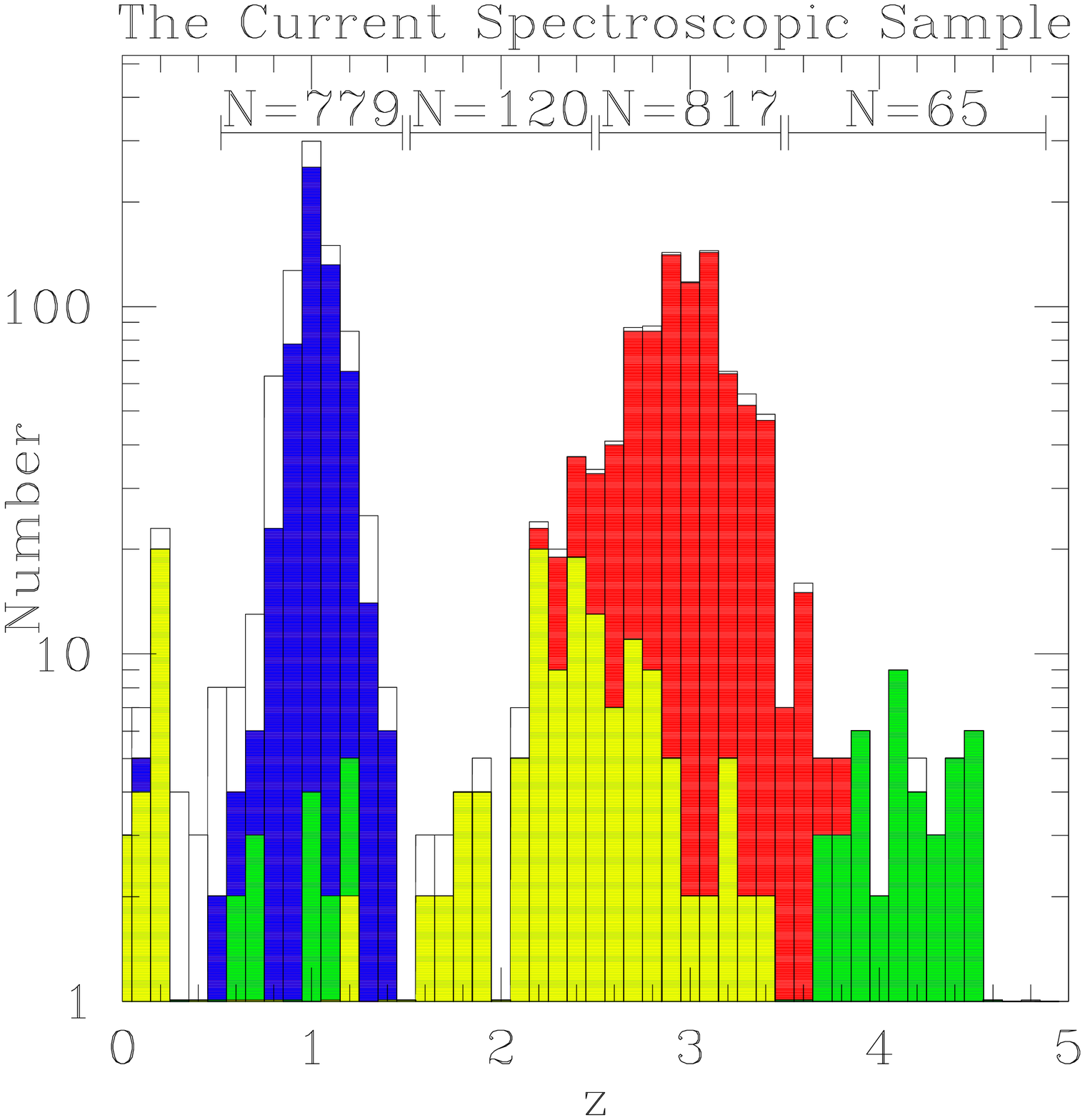}
\caption{
Left:  Examples of color selection
techniques.  The shaded regions are spectra, redshifted
to $z=1$ and $z=3.2$, of a model galaxy that has been forming
stars at a constant rate for 1Gyr.  Absorption from
interstellar and intergalactic hydrogen has been applied
to the spectrum at $z=3.2$.  Superimposed are the
transmissivities of the $U_nG{\cal R}i$ filters.  The
Balmer and Lyman breaks give galaxies at these redshifts
distinctive colors, allowing large numbers of them
to be located in deep images.  These and other
spectral features can of course be used to find galaxies
at other redshifts as well.  Right:  The redshift
histogram of all galaxies in the color-selected
samples of Steidel et al. (1999) and Adelberger et al. (2000).
Different colors correspond to selection criteria aimed
at different redshifts.
}
\end{figure}

Although different strategies for finding galaxies at $z\ga 1$ result
in samples weighted towards different types of objects, there 
is nevertheless significant overlap between the galaxy populations
that are found.  The left panel of Figure 2, showing the Ly-$\alpha$ equivalent width
distribution of color-selected Lyman-break galaxies at $z\sim 3$,
illustrates the point.  About 20\% of Lyman-break galaxies
have equivalent widths large enough to be detected in
standard narrow-band searches for high redshift galaxies.
At fixed continuum luminosity, narrow-band searches detect only
the fraction of galaxies with the largest equivalent widths,
and at fixed equivalent width, color-selected surveys detect only
the fraction of galaxies with the brightest continua; but
the galaxies detected with these techniques appear to belong to
the same underlying population.
Similarly, although the $z\sim 1$ Balmer-break
selection criteria of Adelberger et al. (2000) are designed to
select optically bright star-forming galaxies at $z\simeq 1$,
a substantial fraction of these galaxies (the limited available
data suggests 1 in $\sim 20$, or $\sim 0.2$ per square arcmin to $K_s\simeq 20$)
have the red optical-to-infrared colors ${\cal R}-K_s\ga 5.5$ that
are often thought to be characteristic of extremely dusty or old
galaxies at this redshift.  Many of the same galaxies will therefore
be found both by surveys for old or dusty galaxies at $z\sim 1$ that
exploit their expected large optical-to-infrared colors and
by surveys for star-forming galaxies at $z\sim 1$ that exploit
the Balmer break.
Finally, there is even some overlap between
far-UV selected samples and far-IR selected samples
of galaxies at high redshift, though these two selection
strategies might have been expected {\it a priori}
to find completely different populations of objects.
For example, the two $f_\nu(850\mu{\rm m})>5$mJy sources robustly identified
with star-forming galaxies (as opposed to AGN) at $z\ga 2$,
SMMJ14011+0252 at $z=2.565$ (Ivison et al. 2000) and
West-MMD11 at $z=2.979$ (Chapman et al. 2000), 
have the relatively blue far-UV colors observed in
optically selected galaxies at similar redshifts;
they are typical, aside from their unusually
{\it bright} far-UV luminosities, of
the kind of galaxies found in optical surveys.

The relationship between sub-mm selected
and UV-selected high-redshift populations
can be partially understood with plots
like the right panel of Figure 2, which shows the inferred
distribution of dust opacities among optically
selected galaxies at $z\sim 3$ (Adelberger et al. 2000).
These dust opacities were estimated with a relationship
between $A_{1600}$ and far-UV spectral slope
that is obeyed by starburst galaxies in the
local universe (e.g. Meurer, Heckman, \& Calzetti 1999).
It is not known if high-redshift
galaxies obey this relationship; see \S 4 below.
The majority of galaxies at this redshift
appear to have middling dust opacities and
are therefore far easier to detect in the optical
than at $850\mu$m, but some
galaxies, even in optical surveys, are so
dusty that they would have been easier to detect
with sub-mm rather than optical imaging.
Although relatively dust-free galaxies appear
to dominate high-redshift populations
by number, it is unclear if
they dominate by star-formation rate:  dusty galaxies tend
to have much larger star-formation rates and
this compensates, to some unknown but probably large
extent, for their smaller numbers.  I will discuss this
further in \S 4.

\begin{figure}
\plottwo{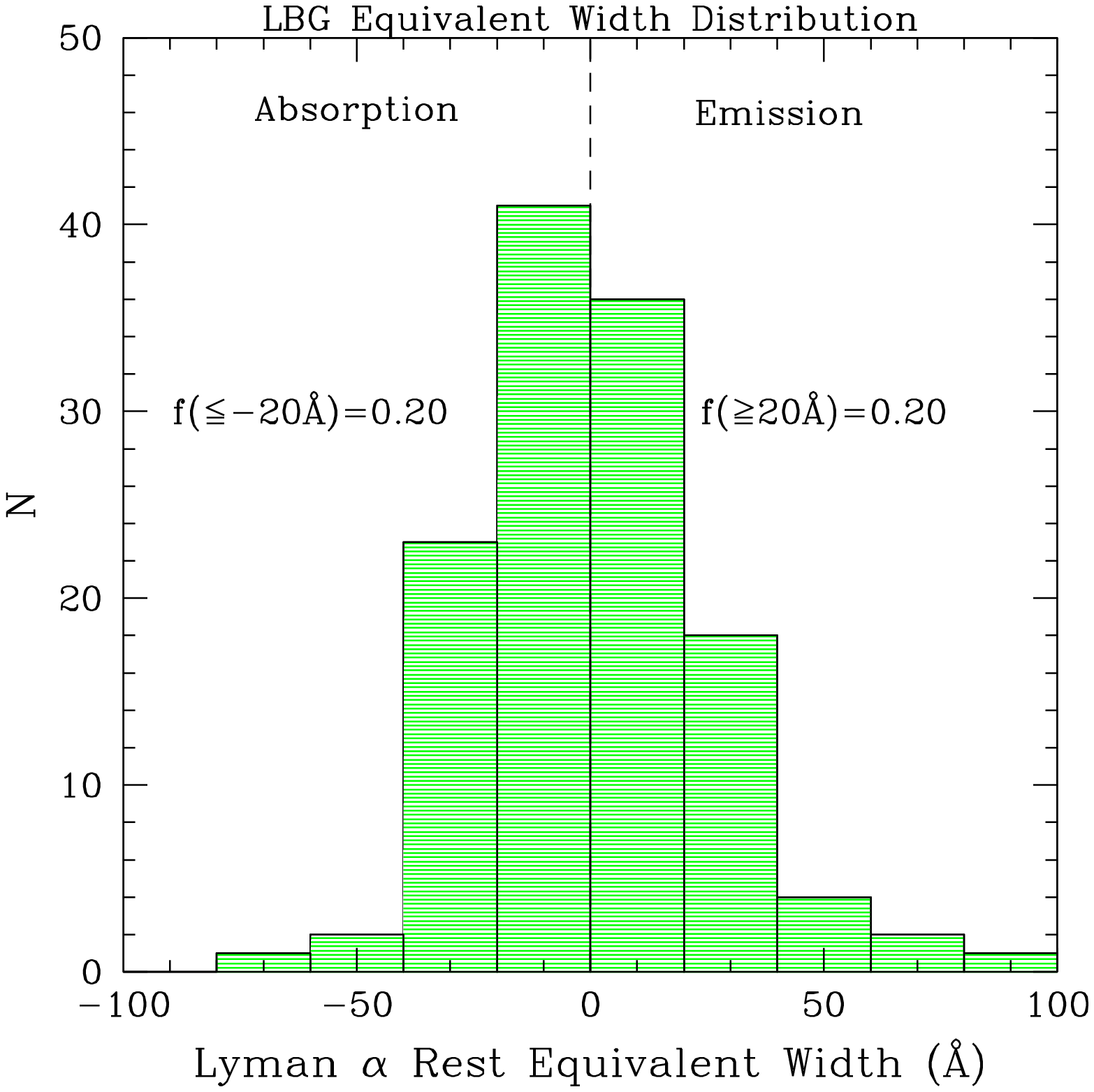}{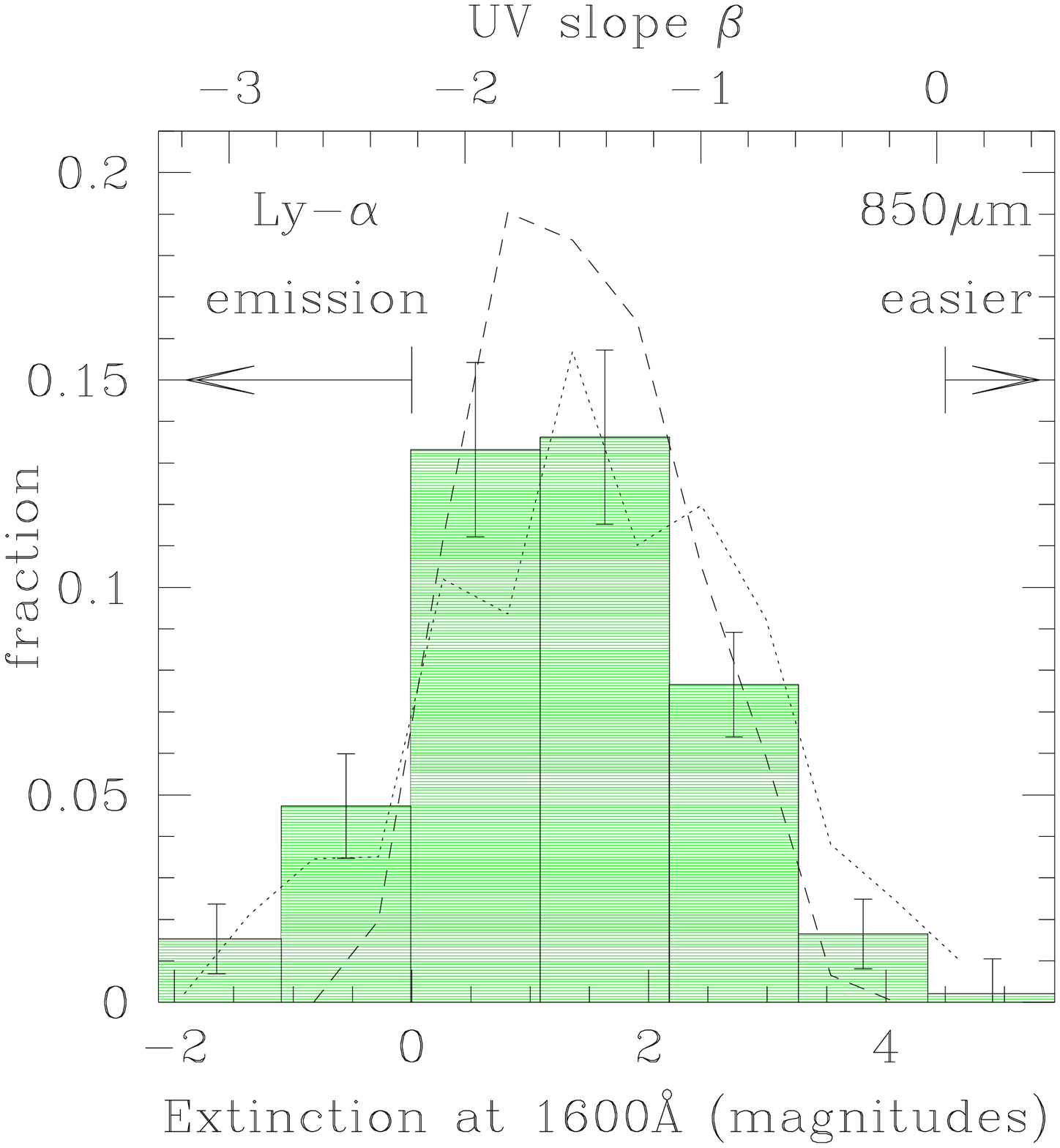}
\caption{
Left:  The equivalent width distribution of Lyman-break galaxies at $z\sim 3$
(Steidel et al. 2000).
Roughly 20\% of LBGs have rest equivalent widths $W_\lambda\ga 20$\AA\
large enough to be detected in standard narrow band surveys.
Right:  The distribution of dust opacities in Lyman-break galaxies
at $z\sim 3$, estimated from their far-UV spectral slopes
$\beta$ ($f_\lambda\propto\lambda^\beta$) as described in the text.  The solid
bars show the best estimate; dotted and dashed lines indicate the
size of systematic uncertainties.  Galaxies
with $A_{1600}<0$ have broad-band colors contaminated
by strong Ly-$\alpha$ emission.  Galaxies with
$A_{1600}\ga 5$ will be too red to satisfy the Lyman-break selection
criteria, but most known galaxies at $z\sim 3$ are much less obscured.
Although relatively unobscured galaxies appear to dominate the
universe at $z\sim 3$ by number, they may not dominate by
star formation; see \S 4.
A galaxy with $f_\nu(850\mu{\rm m})=2$mJy, barely detectable by SCUBA,
will be too faint in the optical to be included in our Lyman-break
survey if it has $A_{1600}\ga 4.5$; a galaxy with ${\cal R}=26$,
barely detectable in the deepest Lyman-break survey fields, will
be too faint at $850\mu$m to be detected by SCUBA if it has
$A_{1600}\la 4.5$; see Adelberger \& Steidel (2000).  
}
\end{figure}

\section{Spatial Clustering in High-Redshift Samples}
The large samples of star-forming galaxies at $z\ga 1$
produced by color-selected surveys allow one to begin
to try to fit star-forming galaxies into the
larger context of structure formation in the universe.
The most obvious way to make a connection between
the observed galaxies and perturbation in the underlying
distribution of matter is to attempt to estimate the
masses associated with individual galaxies by observing
their velocity dispersions.  Unfortunately this 
approach is surprisingly difficult.  To begin with,
it is hard to measure velocity widths for high-redshift
galaxies.  
The [OII], H$\beta$, and [OIII] nebular emission
lines of galaxies at $z\sim 3$, for example, are redshifted
into the bright sky of the near-IR, and
as a result perhaps only 2--3 velocity widths
can be measured per night even with an 8m-class telescope.
A more serious problem is interpreting
the velocity widths that have been measured.
The limited available data suggest that most
Lyman-break galaxies at $z\sim 3$ have
nebular line widths corresponding to $\sigma_{1D}\sim$~70--80~km/s,
for example (Pettini et al. 1998, 2000).  These velocity dispersions are far
smaller than the circular velocities that were
expected for Lyman-break galaxies on a number
of other grounds (e.g. Baugh et al. 1999; Mo, Mao, \& White 1999)
and this has led some
to suggest that Lyman-break galaxies
are low mass ``satellite galaxies'' undergoing
mergers (e.g. Somerville's and Kolatt's contributions
to these proceedings).  But because the baryons
in these galaxies presumably cooled and collapsed
farther than the dark matter before stars began to form, their nebular
line widths are expected to be significantly smaller than
the full circular velocity of the dark matter potential.
The exact size of the difference is not easy to calculate.  Analytic attempts
at the calculation (e.g. Mo, Mao, \& White 1999) rely
on a large number of simplifying assumptions, but
the real situation may not be so simple:
Lehnert \& Heckman (1996) and Kobulnicky \& Gebhardt (2000)
have presented evidence for a complicated relationship
between nebular line widths and circular velocities
in the local universe among late-type and starburst galaxies
that are presumably the closest analogs
to detected high-redshift galaxies.

The spatial distribution of high redshift galaxies
provides an alternate way of making a connection
between star-forming galaxies and perturbations in
the underlying distribution of dark matter.
For a given cosmogony the spatial distribution of
matter at any redshift is straightforward to calculate
with simulations or analytic approximations.
Once large numbers of star-forming galaxies have been
detected near a single redshift we can therefore 
ask, for example, what kinds of collapsed
objects in the expected distribution of mass
at that redshift have the same spatial distribution
as the observed galaxies.  In this way we can attempt
to place star-forming galaxies in the larger
context of structure formation.
Observational constraints on the spatial clustering of galaxies at $z\ga 1$ will be the
main subject of this section, starting at $z\sim 1$ and moving later
to $z\sim 3$.

Attempts to measure
clustering strength for galaxies at $z\sim 1$ have been carried
out by Carlberg et al., Le Fevre et al., and Cohen et al.;
their results are reviewed in Cohen's
contribution to these proceedings.  Here I will focus
instead on previously unpublished results from
a survey of star-forming galaxies at $z\simeq 1.0$
(Adelberger et al. 2000).
This sample consists of several thousand
photometric candidates with ${\cal R}\la 25.5$;
to date redshifts have been obtained for
$\sim 800$ of them.  The mean redshift
of the spectroscopically observed candidates
is $\langle z\rangle\simeq 1.0$ and
the standard deviation is $\sigma_z\simeq 0.10$.
We currently have
uniform spatial spectroscopic sampling
in four $9\arcmin$ and one
$6.5\arcmin$ square fields.  Roughly 100 redshifts
have been obtained in each (Figure 3).

\begin{figure}
\centerline{\epsfysize=3.0truein\epsfbox{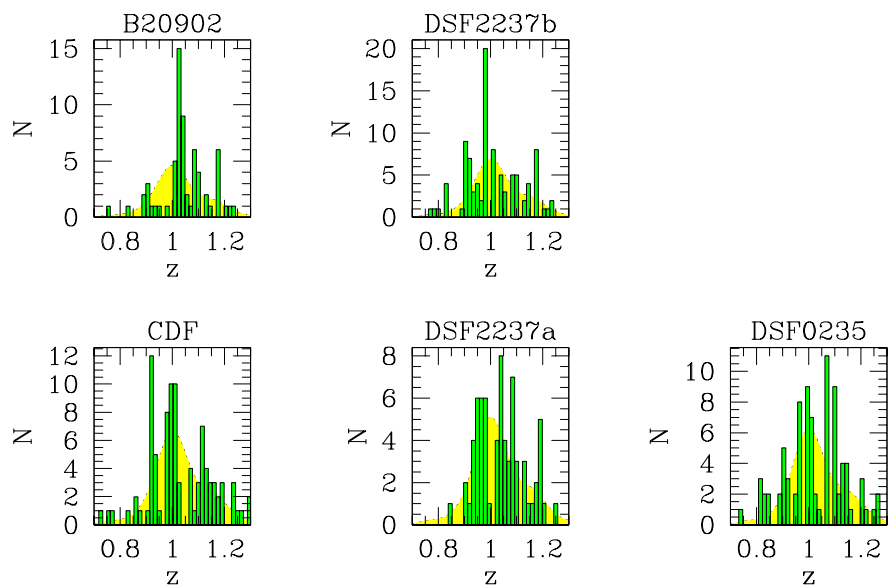}}
\caption{ Redshift histograms in the 5 most complete
fields of the Balmer-break survey at $z\simeq 1$.
The field B20902 is $\sim 6.5\arcmin\times 6.5\arcmin$;
the others are $\sim 9\arcmin\times 9\arcmin$.  The shaded curve
in the background is our estimated selection function, determined
from the full spectroscopic sample.  Only about a quarter
of the photometric candidates were observed spectroscopically
in each field, but the spatial sampling of our spectroscopy
is close to uniform in these fields.
}
\end{figure}

In the four $9\arcmin$ fields, the variance of galaxy counts in cubes of comoving
side length $l\simeq 6 h^{-1}$ Mpc ($\Omega_M=0.3$, $\Omega_\Lambda=0.7$), estimated
as described in Adelberger et al. 1998 from the data in Figure 3, is $\sigma_{\rm gal}^2 = 1.2\pm 0.2$.
For a power-law spatial correlation function of the form $\xi(r)=(r/r_0)^{-1.7}$---which
is consistent with the angular clustering of these galaxies---this
variance corresponds to a comoving correlation length of
$\sim 3 h^{-1}$ Mpc, or a variance of galaxy counts in
spheres with radius $8 h^{-1}$ Mpc of $\sigma_{\rm 8,gal}^2 \sim 0.3$.   
The $8 h^{-1}$ Mpc variance is similar to the expected variance of mass
at $z\simeq 1$ in spheres of the same size, estimated by evolving back
to $z\simeq 1$ with linear theory the value of
$\sigma_8^2$ determined at $z\simeq 0$ from the abundance
of galaxy clusters (e.g. Eke, Cole, \& Frenk 1996).
Balmer-break galaxies are evidently fairly unbiased
tracers of mass fluctuations at $z\sim 1$ (for the $\Omega_M=0.3$,
$\Omega_\Lambda=0.7$ cosmology assumed throughout).
Evolving their clustering strength forward to $z\sim 0$ with the
linear prescription of Tegmark \& Peebles (1998)
suggests that these galaxies are likely the progenitors
of galaxies with $\sigma_8\sim 1$ in the local universe,
i.e. relatively normal galaxies.  This result could
perhaps have been anticipated from the comoving
abundance of Balmer-break galaxies, which, at $\sim 0.02 h^3 {\rm Mpc}^{-3}$ to ${\cal R}\sim 25.5$,
is similar to that of $L_\star$ galaxies in the local universe.

The increase in clustering strength from galaxies in our sample
at $z\simeq 1$ to galaxies at $z\sim 0$ therefore appears
to be relatively easy to understand; it is almost exactly
what one might have expected gravitational instability
to have produced acting on a population of formed
objects.  Remarkably the same is not true at higher redshifts.
Rather than decreasing further, at redshifts $z>1$ the observed
clustering strength of detected star-forming galaxies begins to rise.

Hints that star-forming galaxies at $z\ga 2$ might be
strongly clustered were first provided by targeted surveys
of small and carefully selected volumes, often
around known AGN (e.g. Giavalisco, Steidel, \& Szalay 1994; Le Fevre et al. 1996; Francis et al. 1997;
see also later work by Campos et al. 1999 and Djorgovski et al. 1999).
Further evidence came subsequently from the color-selected
survey of Lyman-break galaxies at $z\sim 3$ (e.g. Steidel et al. 1998).
Figure 4 shows the projected
correlation function $\omega_p(r_p)$ (e.g. Davis \& Peebles 1983)
of galaxies in this sample.
The implied correlation
length, neglecting systematic errors, is $3.8\pm 0.3 h^{-1}$ Mpc 
comoving ($\Omega_M=0.3$, $\Omega_\Lambda=0.7$).  A similar
estimate of the correlation length follows from the relative
variance of Lyman-break galaxy counts in cubes of comoving side-length
$11.4 h^{-1}$ Mpc, $\sigma_{\rm gal}^2 = 0.75\pm 0.25$ (Adelberger et al. 2000;
this value supersedes our previous estimate, which was based on a smaller
data set).

\begin{figure}
\centerline{\epsfysize=3.0truein\epsfbox{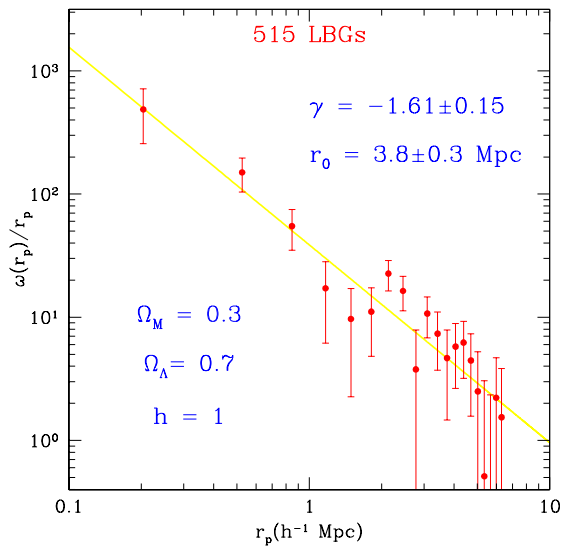}}
\caption{  The projected correlation function of Lyman-break
galaxies at $z\sim 3$, from Adelberger et al. (2000).
The quoted confidence limits on $r_0$ and $\gamma$ do not
include systematic uncertainties, which are likely
at least as large as the random uncertainties.
}
\end{figure}

Gravitational instability acting on a population of galaxies
that had $r_0\sim 4 h^{-1}$ comoving Mpc at $z\sim 3$ would
not produce a population with $r_0\sim 3 h^{-1}$ comoving Mpc
(similar to the observed $r_0$ of Balmer-break galaxies)
at $z\simeq 1$ or a population with $r_0\sim 5 h^{-1}$ comoving
Mpc (similar to normal local galaxies) at $z\simeq 0$.
This can be shown in a crude way by first assuming that
the correlation function of Lyman-break galaxies
selected at $z\sim 3$ would maintain a constant slope
of $\gamma\sim 1.7$ at lower redshifts, and then using
the linear approximation of 
Tegmark \& Peebles (1998) to evolve the
observed clustering of Lyman-break galaxies
at $z\sim 3$ to lower redshifts.
Limited space does not allow a more careful analysis
or the consideration of cosmological
models besides $\Omega_M=0.3$, $\Omega_\Lambda=0.7$,
although both would affect our conclusions somewhat.
In this simplified analysis, we would expect
former Lyman-break galaxies to have $r_0\sim 5 h^{-1}$ comoving
Mpc at $z\simeq 1$ and $r_0\sim 9 h^{-1}$ comoving Mpc
at $z\simeq 0$.

These correlation lengths are significantly larger
than those of Balmer-break galaxies at $z\simeq 1$
and of normal galaxies at $z\simeq 0$, implying
that Lyman-break galaxies are unlikely to be
the progenitors of either population.
What are they instead?  Why is
their spatial clustering so much stronger than
we might naively have expected?
A clue is provided by
their number density, $\sim 4\times 10^{-3} h^3 {\rm Mpc}^{-3}$
($\Omega_M=0.3$, $\Omega_\Lambda=0.7$) to ${\cal R}=25.5$,
which is about 5 times lower
than the number density of normal galaxies in the local
universe or of Balmer-break galaxies with ${\cal R}\la 25.5$
in our $z\simeq 1$ sample.  
Perhaps the relatively rare Lyman-break galaxies are not
progenitors of typical galaxies at $z\la 1$ but
instead are special in some way.  One way in which they
are special is that they are 
the UV-brightest galaxies
(and therefore presumably the most rapidly star-forming galaxies) at $z\sim 3$.
Semi-analytic calculations (e.g. Baugh et al. 1998, Kauffmann et al. 1999)
suggest that the most rapidly star-forming galaxies at high redshift
will reside within the most massive collapsed objects, rather
than within typical collapsed objects, and so perhaps
we can understand the clustering of Lyman-break galaxies
by trying to associate them with massive collapsed objects
at $z\sim 3$ instead of with galaxy populations at lower redshift.

In a classic paper, Kaiser (1984)
showed that in hierarchical models the most massive
collapsed objects at any redshift are more strongly clustered
than the distribution of matter as a whole.
A formalism for estimating the clustering strength
of collapsed objects as a function of mass was subsequently
developed by many authors; see (e.g.) Mo \& White (1996).
Remarkably the observed clustering of Lyman-break
galaxies is indistinguishable (as far as we can tell---see Wechsler's
contribution to these proceedings)
from the predicted clustering of the most massive collapsed
objects at $z\sim 3$ down to a similar abundance (e.g. Adelberger et al. 1998).
This result suggests that there may indeed be a simple
relationship between mass and star-formation rate in high redshift
galaxies, as many semi-analytic models predicted (e.g. Baugh et al. 1998).

If the mass of a galaxy plays a dominant role in determining
its star formation rate, then we might expect the star-formation
rate distribution of $z\sim 3$ galaxies to be related
in a simple way to the distribution of masses of collapsed
objects at $z\sim 3$.  As a crude guess at the star-formation
rate associated with a collapsed ``halo,'' we can take
the mass cooling rate in the halo for large masses
and the mass cooling rate times a number proportional to $v_c^2$ for small masses
where supernova feedback is important (e.g. White \& Frenk 1991).
In this approximation, for cooling dominated by Bremsstrahlung,
we would expect ${\rm SFR}$\lower.5ex\hbox{$\;\buildrel\propto\over\sim\;$}$M^{5/6}$
for large $M$ and ${\rm SFR}$\lower.5ex\hbox{$\;\buildrel\propto\over\sim\;$}$M^{3/2}$
for small $M$.  
Figure 5 shows the slopes of our simplistic ``theoretical'' SFR distribution
in these two limits.  At each abundance the slope of the mass function
was estimated with the Press-Schechter (1974) approximation for
an $\Omega_M=0.3$, $\Omega_\Lambda=0.7$, $h=0.65$, $\sigma_8=0.9$, $\Gamma=0.2$ cosmogony;
changing the values of any of these parameters would change the mass function slope
somewhat.
Also shown in Figure 5 is the observed ``dust-corrected'' luminosity
function of these galaxies, estimated using the $\beta$--$A_{1600}$ correlation
of Meurer et al. (1999) as described in Adelberger \& Steidel (2000).
The observed slopes agree reasonably well with our naive expectations,
but the star-formation rates at any abundance are unexpectedly
high, many times larger than the expected cooling rate for halos
of similar abundance.  (Standard formulae can be used to derive
star-formation rates from the far-UV luminosities shown in 
Figure 5; see, e.g., Madau, Pozzetti, \& Dickinson 1998.)
This has been taken as evidence
that the star formation in Lyman-break galaxies is fueled
not by quiescent cooling but instead by the rapid cooling
that would accompany a merger of two smaller galaxies (e.g.
Somerville's and Kolatt's contributions to these proceedings).
Many uncertain steps lie behind the conclusion that
the star formation rates in Lyman-break galaxies
far exceed their quiescent cooling rates, however.  The star-formation
rates of Lyman-break galaxies could be considerably lower
than is usually deduced, for example, if we are wrong
about the shape of the IMF, about the magnitude of the
required dust corrections, or even about the value of
various cosmological parameters, while the
cooling rates in Lyman-break galaxies could be significantly
higher than usual estimates if we are wrong about
the baryon fraction, the metallicity, or the spatial distribution
of the gas or dark matter in these galaxies.
It would be tremendously interesting if the star formation rates 
of Lyman-break galaxies really were far higher than their quiescent
cooling rates, but I do not think this has been 
conclusively shown.

\begin{figure}
\centerline{\epsfysize=2.0truein\epsfbox{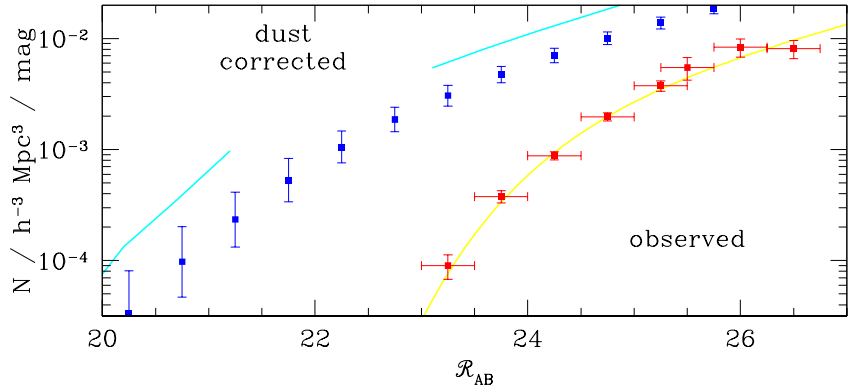}}
\caption{
The observed and dust-corrected luminosity functions of Lyman-break
galaxies at $z\sim 3$.  At this redshift the ${\cal R}$ filter
samples $\sim 1700$\AA\ rest, and so the (dust-corrected)
${\cal R}$ luminosity should be closely correlated with
star-formation rate.  Error bars do not include
some systematic uncertainties; these
uncertainties are especially severe for
${\cal R}_{\rm corrected}\ga 24$---see
Adelberger \& Steidel (2000) before using this plot.  The solid lines show
the expected slope of a naive ``theoretical'' star-formation rate (SFR)
distribution for ${\rm SFR}\propto M^{1.1}$ (left) and ${\rm SFR}\propto M^{1.5}$ (right).
See text.
}
\end{figure}

In any case I will assume for now that Lyman-break galaxies
are strongly clustered because they reside within rare and
massive collapsed objects at $z\sim 3$, and return
to the question of how they might be related to
the Balmer-break galaxies observed at $z\simeq 1$.
The calculation above showed that $z\sim 3$ Lyman-break
galaxies with ${\cal R}\la 25.5$ cannot be the progenitors of
$z\sim 1$ Balmer-break galaxies with ${\cal R}\la 25.5$, and this was hardly
surprising since the abundance of the Lyman-break galaxies
is so much lower.  But suppose we had significantly
deeper $U_nG{\cal R}$ photometry, so that we could
detect fainter Lyman-break galaxies and reach
an abundance similar to that of Balmer-break galaxies
at $z\sim 1$.  Could this deeper population of Lyman-break galaxies
evolve into a population like the Balmer-break galaxies
by $z\sim 1$?  If halo mass and star formation rate
are related in the simple way described above, then
the deep Lyman-break population would be somewhat less strongly
clustered that the current ${\cal R}\la 25.5$ population,
helping to remove the inconsistency between
the observed $r_0$ of Balmer-break galaxies
and the expected $r_0$ of Lyman-break galaxy descendants.
A deep Lyman-break population, with abundance $\sim 15$ times
that of the ${\cal R}\la 25.5$ sample, has in fact been
detected in the HDF, and its correlation length $r_0$
(which cannot be measured very accurately
because of the HDF's small size) appears smaller
than that of the brighter ground-based population (Giavalisco et al. 2000);
but it looks like this effect is not strong enough
to remove the inconsistency in clustering strength
for Lyman-break galaxy descendants and Balmer-break galaxies 
at $z\sim 1$.  We are left with the result that
galaxies selected by the Balmer-break technique at $z\sim 1$
are probably not (for the most part) the descendants of those detected with the Lyman-break
technique at $z\sim 3$.  Because Balmer-break galaxies
appear to be representative of typical star-forming
galaxies at $z\sim 1$, the simplest interpretation
is that Lyman-break galaxies have largely stopped
forming stars by $z\sim 1$---are they perhaps instead
passively evolving into the elliptical galaxies
observed at lower redshifts?

In principle this sort of argument could provide
a very stringent constraint on the star-formation
rates of Lyman-break galaxy descendants at $z\simeq 1$, since
(for $\Omega_M=0.3$, $\Omega_\Lambda=0.7$)
galaxies with as little as $\sim 0.2 h^{-2} M_\odot$/yr of
star formation would have bright enough
UV continua (in the absence of dust obscuration)
to be included in our Balmer-break sample.
But I am still not convinced that the argument
is completely robust;
firmly establishing that former $z\sim 3$ Lyman-break
galaxies are not significantly forming stars at $z\sim 1$
will require a much better understanding than is
currently available of the clustering of
fainter and more numerous Lyman-break galaxies 
at $z\sim 3$ and of what kinds of star-forming
objects might not satisfy our ``Balmer-break'' selection
criteria at $z\simeq 1$.

\section{Dust}
A full observational understanding of star formation at high
redshift can only be achieved if we are able to
detect most of the star formation in representative
portions of the high redshift universe.
But how can we be sure that our surveys are
not missing a large fraction of the star formation
at high redshift?  The sad answer is that we can't.  There are
too many ways that star formation could be hidden from
our surveys for us ever to be sure that we have detected most of it.
Surveys cannot detect the star formation that occurs
in objects below their flux and surface brightness limits, for example,
and they cannot directly detect the formation of the low mass
stars that (at least in the local universe) dominate
the total stellar mass.  At best we can aim for
a reasonably complete census of the formation of massive stars
that occurs in objects above our flux and surface brightness limits.
If these limits are deep enough, and if the high-redshift IMF is
similar enough in all environments to what we have assumed,
then this sort of sample can serve as an acceptable proxy
for a true census of all star formation at high redshift.

What is the best way to produce a reasonably complete survey
of massive star formation?  Massive stars emit most of their
luminosity in the UV, and so naively we might choose
a deep UV-selected survey.
It has become clear in recent years, however, that most
of the UV photons emitted by stars in rapidly
star-forming galaxies are promptly absorbed
by dust, and as a result most of the luminosity
produced by massive stars tends to emerge from these galaxies
in the far-IR where dust radiates.  This is true
in the local universe for a broad range of rapidly
star-forming galaxies, from the famous class of ``Ultra Luminous Infrared
Galaxies'' (ULIRGs, galaxies with $L_{\rm FIR}\ga 10^{12} L_\odot$)
to the much fainter UV-selected starbursts
contained in the IUE Atlas (e.g. Meurer et al. 1999);
and the recent detection of a large extragalactic far-IR  
background (e.g. Fixsen et al. 1998) suggests that it
is likely to have been true at high redshifts as well.

Two implications follow.  The first is that far-IR luminosities
probably provide a better measurement of rapidly star-forming
galaxies' star-formation rates than do UV luminosities.
The second is that even very rapidly star-forming galaxies
may not be detected in a UV selected survey if they
are sufficiently dusty.
Far-IR/sub-mm selected surveys should therefore in principle provide a much better
census of massive star formation at any redshift than UV selected surveys.
$850\mu$m surveys are likely to provide an especially
good census at $1\la z\la 5$, because favorable $K$-corrections
make a galaxy that is forming stars at a given rate appear almost equally bright
at $850\mu$m for any redshift in this range (see Hughes' contribution to these
proceedings), and consequently a flux-limited sample at $850\mu$m is
nearly equivalent to a star-formation limited sample
at $1\la z\la 5$.

This is exactly the kind of sample that detailed
attempts to understand star formation at high redshift
require.  So why then was most of this review devoted
to UV-selected high redshift samples rather than
the apparently superior $850\mu$m samples?
The reason is that $850\mu$m observations are comparatively difficult.
The deepest $850\mu$m image taken with
SCUBA (the current state-of-the-art sub-mm bolometer array)
reached a depth of 2mJy in a $\sim 4$ square arcminute region
of the sky, for example, and five sources were detected (Hughes et al. 1998).
In contrast a modern instrument on a 4m-class optical
telescope can easily obtain photometry to a depth
of $\la 0.1\mu$Jy over a $\sim 500$ square arcminute
region, detecting thousands of galaxies at $z\ga 1$.
Even though optical surveys do not select star-forming
galaxies in the optimal way, most of what we know
about galaxies at high redshift comes (and will continue
to come for many years) from these surveys.
The question---probably the most important question
for those interested in high-redshift galaxy formation---is whether
the large and detailed view of the high redshift universe
provided by UV selected surveys is reasonably complete.
 
This is equivalent to asking whether the galaxies responsible
for producing the $850\mu$m extragalactic background are
bright enough in the rest-frame UV to be included
in current optical surveys.  If they are not, then
optically selected surveys will not be able to 
teach us much of value about high-redshift
star formation despite the wealth of information they contain.

The straightforward way to constrain the UV luminosities
of the objects that produce the $850\mu$m background
is to observe the UV luminosities of known $850\mu$m
sources.  This is difficult in practice, however,
because relatively few $850\mu$m sources have
been detected and each has a significant positional
uncertainty due to SCUBA's large diffraction disk.
Often several optical sources lie within a
sub-mm error box, and a great deal of effort
is required to determine which one is
the true optical counterpart (e.g. Ivison et al. 2000).
Moreover only about 30\% of the $850\mu$m background
can be resolved into discrete sources with current technology,
and so optical observations of detected $850\mu$m sources
cannot conclusively tell us about the optical luminosities
of the galaxies responsible for producing most
of the $850\mu$m background.

An alternate approach, taken by Adelberger \& Steidel (2000), is to
estimate the contribution to the $850\mu$m background from
known optically selected populations at high redshift and
compare this expected background to the observed background
to see if there is significant shortfall.  

Meurer et al. (1999) have shown that the far-IR luminosities of
UV-selected starbursts in the local universe can be estimated to within a factor
of $\sim 2$ from the starbursts' UV luminosities and spectral
slopes $\beta$.  Their $\beta$/far-IR relationship was the
foundation of our calculation.  It is not known if high-redshift
galaxies obey this relationship.  Chapman et al. (2000)
presented evidence that the $850\mu$m fluxes of $z\sim 3$ Lyman-break
galaxies may be somewhat lower than the relationship would suggest,
but the evidence is very marginal when 
the uncertainties in Lyman-break galaxies' predicted
$850\mu$m fluxes are taken into proper account.  Adelberger \& Steidel (2000)
have checked in a number of other ways whether $z\ga 1$ star-forming
galaxies obey the relationship.  Figure 6, showing the predicted
and observed $15\mu$m fluxes of Balmer-break galaxies in the $z\simeq 1$
sample described above, is an example.
The predicted fluxes in Figure 6 assume that these galaxies will follow both
the $\beta$/far-IR correlation of Meurer et al. (1999) and
the correlation between 6--9$\mu$m luminosity
and far-IR luminosity observed in the (star-forming) ULIRG sample of Genzel et al. (1998)
and Rigopoulou et al. (2000).  The $15\mu$m data are from the ISO LW3 observations of Flores et al. (1999).
A large fraction of objects with predicted fluxes below the detection
limit were not detected, and a large fraction of objects with predicted
fluxes above the detection limit were detected.  This plot
therefore provides some support for the notion that
Balmer-break galaxies at $z\simeq 1$
obey the $\beta$/far-IR relationship.  

\begin{figure}
\centerline{\epsfysize=2.0truein\epsfbox{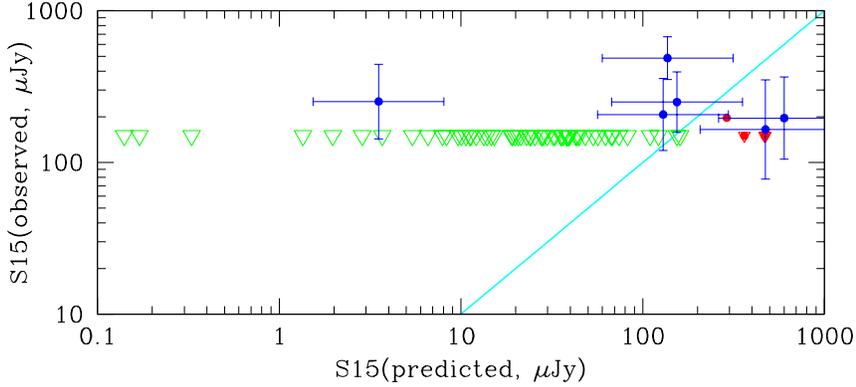}}
\caption{
The predicted and observed $15\mu$m fluxes of Balmer-break
galaxies at $z\sim 1$.  Fluxes were predicted from these galaxies'
UV properties using the $\beta$/far-IR correlation of Meurer et al. (1999)
and the 6--9$\mu$m/far-IR correlation observed among rapidly star-forming
galaxies in the local universe.  The majority of these galaxies
are undetected at a $3\sigma$ limit of $150\mu$Jy;
uncertainties on the predicted fluxes of these galaxies have
been omitted for clarity, and the upper limits to their
observed fluxes are indicated with downward pointing triangles.
The six detections are indicated by points with
error bars in both $x$ and $y$; the one
with $S_{15,{\rm predicted}}\ll S_{15,{\rm observed}}$
has a significant separation between its optical
and mid-IR centroids and may be a misidentification.
}
\end{figure}

If we assume that all optically selected galaxies at $z\ga 1$ obey this
relationship, then we can make crude estimate of their total
contribution to the $850\mu$m background.  In this calculation
I will assume further that the comoving star formation density in
optically selected populations is constant for $1<z<5$ (e.g. Steidel et al. 1999),
that the dust SEDs of optically selected galaxies at $1<z<5$ are similar
to those of starbursts and ULIRGs in the local universe,
and that the (unknown) luminosity and $\beta$ distributions of optically
selected galaxies at $1<z<5$ are similar to those measured for
Lyman-break galaxies at $z\sim 3$.  (See Adelberger \& Steidel 2000
for a more complete discussion.)  Under these assumptions
optically selected populations at $1<z<5$ would be expected
to produce $850\mu$m number counts and background
that are surprisingly close to the observations (Figure 7).

Although the overall agreement is good, within the substantial
uncertainties, there appear to be significant differences
at the brightest $850\mu$m fluxes.  Optically
selected galaxies cannot easily (it seems) account for
the large number of observed sources with $f_\nu(850\mu{\rm m})\ga 6$mJy,
and indeed Barger, Cowie, \& Richards (1999) have shown
that these sources tend to have extremely faint optical counterparts.
It is possible that these bright sources are associated with AGN,
rather than the star-forming galaxies we have included in our
calculation, but in any case it appears that the bulk of the
$850\mu$m background could have been produced by known
optically selected populations at high redshift.
This claim rests on a large number of assumptions that could
easily be wrong, but I am willing to bet \$100
that when ALMA finally resolves the $\sim{\rm mm}$ background
we will discover that most of it is produced by galaxy populations
already detected and studied in the rest-frame UV.
Any takers?

\begin{figure}
\centerline{\epsfysize=3.0truein\epsfbox{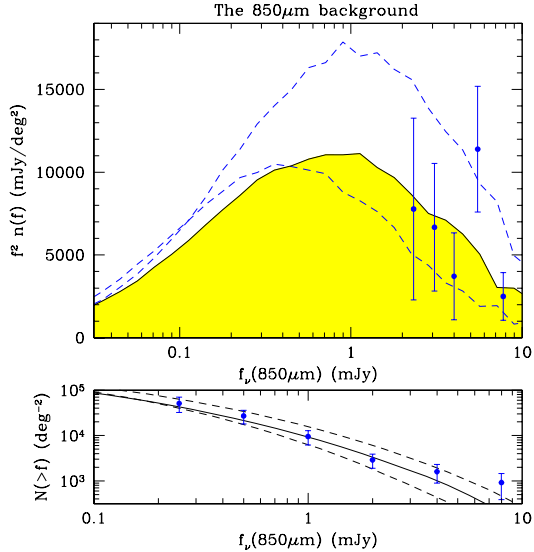}}
\caption{
Top panel:  The contribution to the $850\mu$m background from
UV-selected populations, calculated as described
in the text.  Constraints
on the bright end from Barger, Cowie, \& Sanders (1999) are shown as points
with uncertainties.  Our best guess
at these galaxies' background contribution (the area of the shaded curve)
agrees to within 10\% of the measured value from Fixsen et al. (1998),
but the excellent agreement is likely only a coincidence; the
systematic uncertainties in this calculation are very large.
The dashed lines illustrate the systematic uncertainty due to
possible errors in our derived $\beta$ distribution alone.
Bottom panel:  Observed and predicted $850\mu$m cumulative number counts.
The observed number counts, from Blain et al. 1999, are denoted by
points with error bars.  Our best guess prediction is the solid line;
dashed lines are as above.
}
\end{figure}

\acknowledgments
I would like to thank the organizers for support and
for considerable patience.
Special thanks are due to my collaborators
C. Steidel, M. Dickinson, M. Giavalisco, M. Pettini, and A. Shapley
for their contributions to this work.  Super special thanks
go to C.S. for his comments on an earlier draft.


\end{document}